\documentclass[twocolumn]{jpsj3} 
\usepackage{txfonts}
\usepackage{color}
\usepackage{float}

\title{Microscopic Evidence of Direct Coupling between Magnetic and Superconducting Order Parameters in BaFe$_2$(As$_{1-x}$P$_x$)$_2$}

\author{Tetsuya \surname{Iye}$^{1}$\thanks{E-mail address: tiye@scphys.kyoto-u.ac.jp},
\name{Yusuke \surname{Nakai}}$^{1}$\thanks{Present address: Department of Physics, Graduate School of Science and Engineering, Tokyo Metropolitan University, Hachioji, Tokyo 192-0397, Japan},
\name{Shunsaku \surname{Kitagawa}}$^{1}$,
\name{Kenji \surname{Ishida}}$^{1}$,
\name{Shigeru \surname{Kasahara}}$^{2}$,
\name{Takasada \surname{Shibauchi}},
\name{Yuji \surname{Matsuda}},
and
\name{Takahito \surname{Terashima}}$^{2}$
}
\inst{\address{Department of Physics, Graduate School of Science, Kyoto University, Kyoto 606-8502, Japan} \\
$^{1}$\address{Transformative Research Project on Iron Pnictides (TRIP), Japan Science and Technology Agency (JST), Tokyo 102-0075, Japan}\\
$^{2}$\address{Research Center for Low Temperature and Materials Sciences, Kyoto University, Kyoto 606-8502, Japan }\\
}
\abst{The coexistence of magnetism and superconductivity in the isovalent-P-substituted BaFe$_2$(As$_{1-x}$P$_x$)$_2$ has been investigated microscopically by $^{31}$P-NMR measurements.
We found that superconducting (SC) transition occurs in a magnetic region with static ordered moments and that the moments are reduced below the SC transition temperature $T_{\rm c}$ in the samples near the phase boundary of magnetism and superconductivity.
Our results indicate that magnetism and superconductivity coexist spatially but compete with each other on the same Fermi surfaces.
The coexistence state is qualitatively different from that observed in other unconventional superconductors and gives a strict constraint on the theoretical model for superconductivity in BaFe$_2$(As$_{1-x}$P$_x$)$_2$.}

\kword{Iron-pnictide superconductors, NMR, coexistence between magnetism and superconductivity}

\begin{document}
\maketitle
The newly discovered iron-pnictide high-$T_{\rm c}$ superconductivity appears where antiferromagnetism is suppressed by chemical substitution or pressure\cite{KamiharaJACS08,IshidaJPSJ09,PaglioneNP10,JohnstonAP10}.
Within various iron-pnictide superconductors, we have focused on the isovalent-P-substituted BaFe$_2$(As$_{1-x}$P$_x$)$_2$ system\cite{Jiang,KasaharaPRB10} and carried out comprehensive NMR studies on the system. 
This is because the isovalent P substitution introduces a negligible change in carrier concentration but changes magnetic interactions systematically.
In addition, very clean single crystals are available for various measurements\cite{ShishidoPRL10}.
By investigating the spin dynamics probed via $^{31}$P-NMR as a function of P concentration, we showed that the maximum $T_{\rm c}$ is observed at the P concentration at which magnetic order vanishes; thus, we suggested that the antiferromagnetic (AFM) fluctuations associated with quantum criticality play a central role in obtaining high-$T_{\rm c}$ superconductivity\cite{KasaharaPRB10,ShishidoPRL10,NakaiPRL10,IshidaPhysique11}.

We consider that the study of the coexistence of magnetism and superconductivity, which is observed in the phase boundary, would give important information about the superconducting (SC) properties and mechanism. 
There are experimental reports not only from macroscopic and neutron scattering measurements\cite{PrattPRL09,NandiPRL10,AvciPRB11} but from microscopic measurements\cite{DrewNM09,LaplacePRB09,JulienEPL09,Wiesenmayer11} of the coexistence and competition of magnetism and superconductivity in iron pnictides.
We point out that diffraction measurements give magnetic moments averaged over a wider spatial region and cannot provide microscopic information about the coexistence.
It remains an open question whether an SC phase transition occurs in a magnetic region of a sample or the superconductivity in a magnetic region is induced by the penetration of a nonmagnetic SC region and whether the same or different electronic states contribute to magnetism and superconductivity. 
In this paper, we report the occurrence of the SC transition in a magnetic region of the spin-density-wave (SDW) ordering, which is concluded from the investigation of internal field and nuclear spin-lattice relaxation rate $1/T_1$ by site-selective NMR measurements. 

We used BaFe$_2$(As$_{0.80}$P$_{0.20}$)$_2$ and BaFe$_2$(As$_{0.75}$P$_{0.25}$)$_2$, which are located near the phase boundary as shown in Fig. 1(a). 
We used a collection of single crystals ($\sim$100 mg) for our NMR measurements. The preparation of the single-crystal samples is described in the literature\cite{KasaharaPRB10}. 
The P concentration values were determined using an energy-dispersive X-ray analyzer and confirmed by lattice-constant measurements since they follow Vegard's law.
Figures 1(b) and 1(c) show the temperature dependence of the resistivity $\rho_{xx}$ values of the $x$ = 0.20 and 0.25 samples. 
The structural and magnetic transitions were determined from the anomalies in the temperature derivative of $\rho_{xx}$ and are plotted in Fig. 1(a).
The magnetic transition temperature derived from $\rho_{xx}$ is in good agreement with the NMR results shown later.
Each single crystal was fixed randomly with a GE varnish to prevent preferential alignment in the applied magnetic field due to the anisotropy of magnetization and the SC Meissner effect.
The ac susceptibility measurements indicate nonbulk superconductivity at ambient pressure but bulk superconductivity at $T_{\rm c}^* \sim 13$ K at $P \sim 2$ GPa in the $x = 0.20$ sample [Fig. 1(d)].
In the $x = 0.25$ sample, the gradual development of a Meissner signal below $T_{\rm c}^{\rm on} \sim 30$ K and the SC transition suggested by a broad peak at around $T_{\rm c}^* \sim 14$ K in the temperature derivative of the Meissner signal are observed [Fig. 1(e)]. 
Note that the Meissner signal saturates at low temperatures, indicative of bulk superconductivity in these samples. 
The NMR spectrum becomes broader and inhomogeneous below the magnetically ordered temperature $T_{\rm N}$.
To detect a site-selective electronic state in the inhomogeneous state, we measured the $^{31}$P nuclear spin-lattice relaxation rate $1/T_1$ across the resonance signal.
We applied as small RF fields as possible for the site-selective $1/T_1$ measurements in order to obtain the recovery curve of nuclear magnetization with a single $T_{1}$ component.

\begin{figure}[htbp]
\begin{center}
\includegraphics[width=75mm]{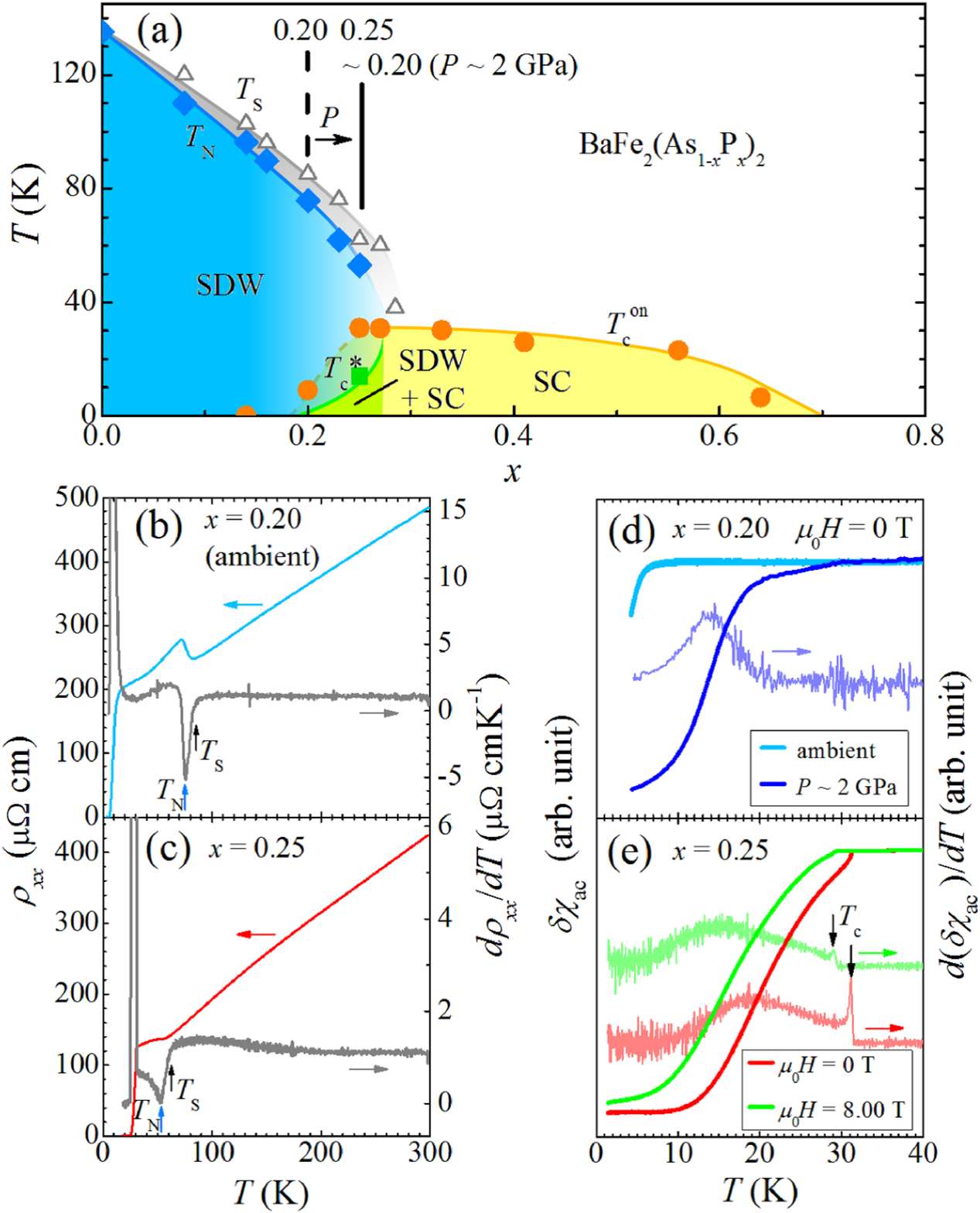}
\end{center}
\caption{(Color online) (a) Phase diagram of BaFe$_2$(As$_{1-x}$P$_x$)$_2$.
Open triangles and closed diamonds show the structural transition temperature $T_{\rm S}$ and magnetic transition temperature $T_{\rm N}$, respectively.
The onset of superconductivity $T_{\rm c}^{\rm on}$ determined by ac susceptibility measurements is indicated by a closed circle.
$T_{\rm c}^{*}$ at $x=0.25$ is denoted by a closed square (see text).
(b) and (c) $\rho_{xx}(T)$ and the temperature derivative of the $\rho_{xx}(T)$ in (b) $x$ = 0.20 and (c) $x$ = 0.25 at ambient pressure.
The structural and magnetic transitions are determined using $d\rho_{xx}/dT$, as shown by arrows.
(d) and (e) Temperature dependence and the temperature derivative of the Meissner signal at (d) ambient pressure and $P \sim$ 2 GPa of $x$ = 0.20, and at (e) ambient pressure of $x$ = 0.25.}
\end{figure}

Figure 2(a) shows the temperature variation of the $^{31}$P-NMR spectrum for $x = 0.25$, which was obtained by sweeping magnetic fields.
A single sharp spectrum was observed above $T_{\rm N}$, but no anomaly was detected in the NMR spectrum at the structural transition $T_{\rm S}$ determined by $\rho_{xx}$.
Below $T_{\rm N}$, a broad NMR spectrum with a Gaussian shape develops gradually and coexists with a sharp peak at around $T_{\rm c}^{\rm on} \sim 30$ K.
We measured $1/T_1$ at the sharp and broad peaks shown by the solid black and dashed red arrows, respectively.
Figure 2(c) shows the temperature dependence of $(T_{1}T)^{-1}$ at those peaks.
$(T_{1}T)^{-1}$ measured at the sharp peak continues to increase down to $T_{\rm c}$, indicative of the absence of a magnetic order at $T_{\rm N}$, but $(T_{1}T)^{-1}$ at the broad peak shows a kink at $T_{\rm N}$, indicative of the presence of a magnetic order.
These results suggest that the sharp (broad) NMR peak arises from the region where P concentration is slightly higher (lower) than $x = 0.25$. 
Thus, the two-peak structure around $T_{\rm c}^{\rm on}$ is ascribed to the spatial distribution of the P concentration, since the variation in $T_{\rm N}$ is very steep against the P concentration, as seen in Fig. 1(a).

The $^{31}$P-NMR spectrum in the AFM state gives information about the ordered moments, since the internal field at the P site originates from the moments. 
The shape of the spectrum reflects the field distribution at the P site; thus, it is related to the magnetic structure of the moments. 
The average of the internal field is related to the magnitude of the ordered moment, which is the order parameter of the AFM state. 
Since we used randomly oriented samples for the measurements and the nuclear spin of $^{31}$P is $I = 1/2$, a rectangular NMR spectrum is expected when the ordered moments are in a commensurate structure. 
Such a rectangular NMR spectrum was actually observed in slightly P-substituted samples, such as the $x$ = 0.07 sample \cite{Iye}.
The broad Gaussian-shaped NMR spectrum indicates the distribution of the internal field at the P site and suggests an incommensurate spin structure. 
The magnetic structure changes from a commensurate structure to an incommensurate structure (SDW type) with the P substitution.
The averaged internal field $\langle H_{\rm int}\rangle$ at the P site is coupled with the ordered moment $\langle M\rangle$ as $\langle H_{\rm int}\rangle$ = $^{31}A_{\rm hf} \langle M\rangle$ with the hyperfine coupling constant $^{31}A_{\rm hf} \sim 0.32$ T/$\mu_{\rm B}$\cite{hyperfine}.
$\langle H_{\rm int}\rangle$ is evaluated from the increase in the second moment $\sqrt{\left<\Delta H^2\right>}$ of each NMR spectrum below $T_{\rm N}$ $\left( \langle H_{\rm int}(T) \rangle \equiv \sqrt{\left<\Delta H(T<T_{\rm N})^2\right>} - \sqrt{\left<\Delta H(T_{\rm N})^2\right>}\right)$, and $\sqrt{\left<\Delta H^2\right>}$ is determined as
\begin{equation}
\sqrt{\left<\Delta H^2\right>}=\left[\frac{\int_0^{\infty}\left(H-H_{\rm av}\right)^2g(H)dH}{\int_0^{\infty}g(H)dH}\right]^{1/2},
\end{equation}
where $g(H)$ denotes the NMR intensity against magnetic fields and $H_{\rm av}$ is the center of gravity of the NMR spectrum. 
Figure 2(b) shows the temperature variations in $\langle H_{\rm int} \rangle$ and estimated $\langle M \rangle$ using $^{31}A_{\rm hf}$.
$\langle H_{\rm int} \rangle$ increases below $T_{\rm N}$ but decreases below $T_{\rm c}^* \sim 14$ K. 
$(T_{1}T)^{-1}$ measured at the magnetic broad signal decreases clearly below $\sim 14$ K, as shown in Fig. 2(c), indicative of the occurrence of superconductivity in the magnetic region of the sample. 
In addition, we measured $(T_{1}T)^{-1}$ across the $^{31}$P-NMR spectra at 20 K ($>T_{\rm c}^*$) and 5 K ($<T_{\rm c}^*$), as shown in Figs. 2(d) and 2(e), and confirmed that $(T_{1}T)^{-1}$ in all regions of the spectrum at 5 K is smaller than that at 20 K. 
This indicates that superconductivity occurs over the entire region of the sample. 
The sharp decrease in $(T_{1}T)^{-1}$ at $T_{\rm c}^*$ means that superconductivity is a phase transition in the magnetic region and excludes the possibility that superconductivity in the magnetic region is induced by the penetration of the SC region in the sample. 

A similar nature of coexistence of magnetism and superconductivity was also observed in the $x$ = 0.20 sample at a pressure of $\sim$ 2 GPa.
\begin{figure*}[t]
\begin{center}
\includegraphics[width=14cm]{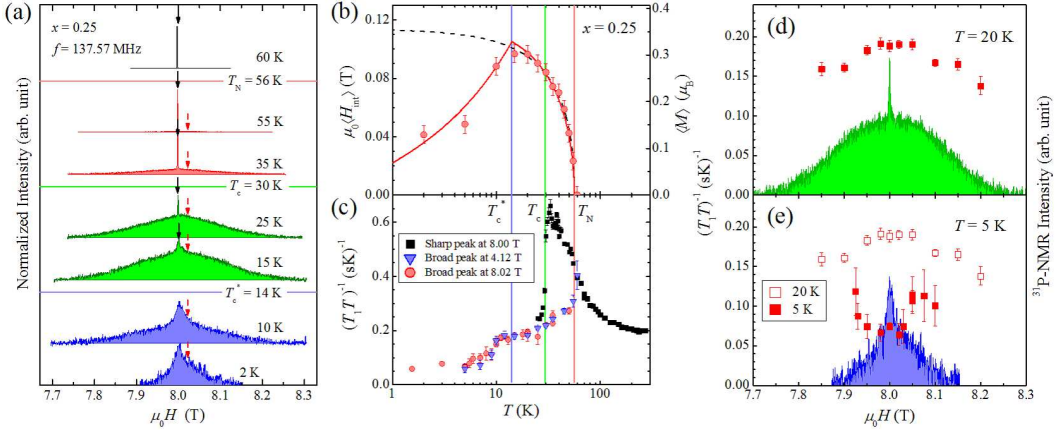}
\end{center}
\caption{(Color online) (a) Field-swept $^{31}$P-NMR spectra in BaFe$_2$(As$_{0.75}$P$_{0.25}$)$_2$.
The solid black and dashed red arrows indicate the magnetic fields where the $1/T_1$ of the paramagnetic and AFM states is measured.
(b) Temperature dependence of the averaged internal field $\langle H_{\rm int} \rangle$ estimated from the second moment of the observed NMR signals.
$\langle H_{\rm int} \rangle$ increases below $T_{\rm N}$ but decreases at $T_{\rm c}^*$.
The broken line indicates the fitting of the data ranging from $T_{\rm c}^*$ to $T_{\rm N}$ to the phenomenological formula of $c(T_N-T)^{\beta}$ with $c$ = 0.02, $T_{\rm N}$ = 56.2 K, and $\beta$ = 0.42.
The red solid curve is the fit with the GL model in the case of homogeneous coexistence\cite{FernandesPRB8110,FernandesPRB8210} [$(M(T)/M_0)^2 = A(T_{\rm N}-T) + B(T-T_{\rm c}^*)$ with $M_0 = 0.35$ $\mu_{\rm B}$, $A = 2.03 \times10^{-2}$ K$^{-1}$, and $B = 8.33 \times 10^{-2}$ K$^{-1.}$].
(c) Temperature dependence of $(T_{1}T)^{-1}$ measured at the sharp paramagnetic NMR signal (closed squares) and the broad magnetic NMR signal (triangles and circles).
(d) and (e) $(T_{1}T)^{-1}$ measured across the $^{31}$P spectra at (d) $T$ = 20 K ($>T_{\rm c}^*$) and (e) $T$ = 5 K ($<T_{\rm c}^*$).
The $(T_{1}T)^{-1}$ over the entire spectrum at 5 K is smaller than that at 20 K, indicative of the occurrence of superconductivity over the entire region of the sample.  }
\end{figure*}
Figures 3(a) and 3(b) show the field-swept NMR spectra and the temperature dependence of the $\langle H_{\rm int} \rangle$ at the P site and estimated $\langle M \rangle$, which are determined by the same procedure as that in the case of the $x$ = 0.25 sample.
For comparison,  $\langle H_{\rm int} \rangle$ at ambient pressure is also shown in Fig. 3(b).
$\langle H_{\rm int} \rangle$ increases below $T_{\rm N} \sim 73$ K ($60$ K) at ambient pressure ($P\sim 2$ GPa) but decreases below $T_{\rm c}^* \sim$ 13 K at $P \sim 2$ GPa, as observed in the $x$ = 0.25 sample. 
It should be noted that the paramagnetic signal was not observed at all in the sample, indicating that the entire region of the $x$ = 0.20 sample at $P \sim 2$ GPa is in the magnetic state below $T_{\rm N}$ and then shows superconductivity below $T_{\rm c}^*$.
These results indicate the microscopic coexistence of magnetism and superconductivity, and the suppression of the magnetic order parameter below $T_{\rm c}^*$ both for the $x$ = 0.25 sample at ambient pressure and for the $x$ = 0.20 sample at $P \sim$ 2 GPa.

Now, we compare the present results with experimental results in other iron pnictides and unconventional superconductors. 
It was revealed by muon spin rotation ($\mu$SR) that strongly disordered but static magnetism coexists with superconductivity in the narrow range of $0.1 < x < 0.13$ on SmFeAsO$_{1-x}$F$_x$ with the ``1111'' structure\cite{DrewNM09}. 
However, neither the suppression of the ordered moments below $T_{\rm c}$ nor the SC transition in a magnetic region was observed in SmFeAsO$_{1-x}$F$_x$.
The suppression of the ordered moments below $T_{\rm c}$, which is observed in BaFe$_2$(As$_{1-x}$P$_x$)$_2$, was also reported in Ba(Fe$_{1-x}$Co$_x$)$_2$As$_2$ by neutron scattering measurements \cite{PrattPRL09} and in Ba$_{1-x}$K$_x$Fe$_2$As$_2$ by $\mu$SR  measurements \cite{Wiesenmayer11}.
The competition between the magnetic and SC order parameters as well as the spatial coexistence are common features of the ``122'' compounds. 
It thus seems that the relationship between the magnetic and SC phases is different between the ``1111'' and ``122'' compounds.
The spatial coexistence is also realized in heavy-fermion superconductors, whereas the nature of coexistence differs from that in BaFe$_2$(As$_{1-x}$P$_x$)$_2$, i.e., magnetism and superconductivity occur at different parts of the Fermi surfaces.
For example, it was reported that the spatial coexistence is observed microscopically in CeCo(In$_{1-x}$Cd$_x$)$_5$\cite{NairPNAS10,UrbanoPRL07} and CeRhIn$_5$ at a certain pressure\cite{KawasakiPRL03,YashimaPRB09}, where $T_{\rm N}$ is higher than $T_{\rm c}$.
In these compounds, the magnetically ordered moments develop below $T_{\rm N}$; however, they become constant below $T_{\rm c}$ without further change at low temperatures.
This suggests that magnetism and superconductivity emerge from different parts of the Fermi surfaces, thus a single 4$f$ state could satisfy the coexistence of magnetism and superconductivity.
The suppression of the ordered moments below $T_{\rm c}$ was not observed in the U-based superconductor UPd$_2$Al$_3$ or UNi$_2$Al$_3$, where plural 5$f$ electrons are subdivided into localized and itinerant electrons\cite{KyogakuJPSJ93,MatsudaPRB97,SatoNature01}.
This is also the case in ferromagnetic superconductors, such as UGe$_2$\cite{HuxleyPRB01}, URhGe\cite{AokiNature01}, and UCoGe\cite{deVisserPRL09,OhtaJPSJ10}.
In contrast, the suppression of the ordered moments below $T_{\rm c}$ can be well understood by the assumption that the same Fermi surfaces contribute to both magnetic ordering and superconductivity with a competitive relationship\cite{FernandesPRB8210,VorontsovPRB10,CvetkovicPRB09}.
In BaFe$_2$(As$_{1-x}$P$_x$)$_2$, cylindrical Fermi surfaces, particularly hole Fermi surfaces, are modified to be more three-dimensional, and nesting between hole and electron Fermi surfaces becomes weaker with P substitution, resulting in the suppression of the AFM order.
It is considered that the nesting vector becomes distributed and the AFM order (superconductivity) occurs at stronger (weaker) nesting vectors.
In this regard, the competitive relationship between the two phases might have a common aspect with that in cuprate superconductors\cite{LakeNature02}, whose AFM order is induced by the applied strong magnetic field that imposes the vortex lattice. 

The competition between superconductivity and AFM ordering near the phase boundary can be described in terms of a Ginzburg-Landau (GL) theory of coupled order parameters\cite{KatoPRB88,ZhangScience,FernandesPRB8210,VorontsovPRB10}. 
The homogeneous part of the free energy is given by
\begin{eqnarray*}
\lefteqn{F_{\rm GL}(\Delta, \mbox{\boldmath $M$})} \\
& = &\int{dr\left(\frac{a_s}{2}|\Delta|^2+\frac{u_s}{4}|\Delta|^4+\frac{\gamma}{2}|\Delta|^2\mbox{\boldmath $M$}^2+\frac{a_m}{2}\mbox{\boldmath$M$}^2+\frac{u_m}{4}\mbox{\boldmath$M$}^4\right)},
\end{eqnarray*}
where $\Delta$ and $\mbox{\boldmath $M$}$ denote the SC and AFM order parameters, respectively. 
The leading term in the order-parameter competition is characterized by the coefficient $\gamma$, which determines the phase diagram and the character of two phase transitions; SC and AFM states coexist homogeneously with two second-order phase lines for $0<\gamma<\sqrt{u_su_m}$, but two phases compete and are separated by a first-order transition for $\gamma>\sqrt{u_su_m}$. 
Our experimental results, indicating the homogeneous coexistence and presence of the AFM and SC transitions at different temperatures, conclude that the former ($0<\gamma<\sqrt{u_su_m}$) condition is realized in BaFe$_2$(As$_{1-x}$P$_x$)$_2$. 
According to the recent theoretical studies, an extended $s$-wave ($s_{\pm}$-wave) state\cite{FernandesPRB8210,VorontsovPRB10}, in which the SC gap changes sign between the hole and electron bands, satisfies the $0<\gamma<\sqrt{u_su_m}$ condition and can coexist with an incommensurate SDW state over a much wider range of parameters than a conventional $s$-wave state.
In fact, the temperature dependence of $\langle M \rangle$ can be fit with the GL expression in the case of homogeneous coexistence\cite{FernandesPRB8110,FernandesPRB8210}, as shown in Figs. 2(b) and 3(b) with red curves. 
In this sense, our results are consistent with the $s_{\pm}$-wave state.

In conclusion, our NMR results on BaFe$_2$(As$_{1-x}$P$_x$)$_2$ reveal the spatial coexistence of AFM and SC states, and the direct coupling between the two order parameters.
The coexistence behavior is consistently interpreted by the $s_{\pm}$ model\cite{FernandesPRB8210}.
As far as we know, iron-pnictide superconductors with the ``122" structure are the first examples that magnetic ordered moments are significantly suppressed by the occurrence of superconductivity, although such coexistence has been expected from theoretical studies\cite{KatoPRB88}.
Our results strongly suggest that antiferromagnetism and superconductivity occur at the same Fermi surfaces and compete with each other.
\begin{figure}[t]
\begin{center}
\includegraphics[width=\hsize]{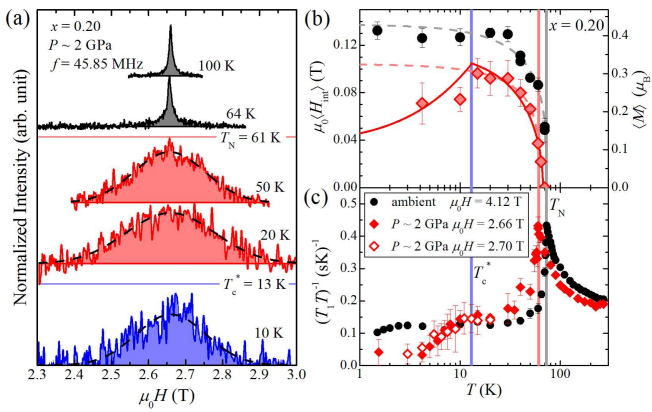}
\end{center}
\caption{(Color online) (a) Field-swept $^{31}$P-NMR spectra in BaFe$_2$(As$_{0.80}$P$_{0.20}$)$_2$ at $P \sim 2$ GPa.
(b) Temperature dependence of the averaged internal field $\langle H_{\rm int} \rangle$ estimated from the second moment of the observed NMR signals at ambient pressure (black dots) and at $P \sim 2$ GPa (diamonds).
$\langle H_{\rm int} \rangle$ increases below $T_{\rm N}$ but decreases below $T_{\rm c}^*$ at $P \sim 2$ GPa.
The broken lines indicate the fitting of the data ranging from $T_{\rm c}^*$ to $T_{\rm N}$ to the phenomenological formula of $c(T_{\rm N}-T)^{\beta}$ with $c$ = 0.04 (0.04), $T_{\rm N}$ = 73 (61.0) K, and $\beta$ = 0.29 (0.25) for $P = 0$ (2) GPa.
The red curve is the fit with the GL model in the case of homogeneous coexistence\cite{FernandesPRB8110,FernandesPRB8210} [$(M(T)/M_0)^2 = A(T_{\rm N}-T) + B(T-T_{\rm c}^*)$ with $M_0 = 0.33$ $\mu_{\rm B}$, $A = 1.87 \times 10^{-2}$ K$^{-1}$, and $B = 8.74 \times 10^{-2}$ K$^{-1}$].
(c) Temperature dependence of $(T_{1}T)^{-1}$ values measured in $\mu_0 H$ = 4.12 T at ambient pressure and in $\mu_0 H \sim 2.7$ T at $P \sim 2$ GPa.  }
\end{figure}

\begin{acknowledgment}
The authors thank S. Yonezawa and Y. Maeno for experimental support and valuable discussions, and R. Ikeda, H. Ikeda, and R. M. Fernandes for valuable discussions.
This work was partially supported by the Kyoto University LTM Center, the ``Heavy Electrons" Grant-in-Aid for Scientific Research on Innovative Areas (No. 20102006) from the Ministry of Education, Culture, Sports, Science, and Technology (MEXT) of Japan, a Grant-in-Aid for the Global COE Program ``The Next Generation of Physics, Spun from Universality and Emergence'' from MEXT of Japan, and a Grant-in-Aid for Scientific Research from JSPS, KAKENHI (S) (No. 20224015) and (A) (No. 23244075).


\end{acknowledgment}

\end{document}